\documentclass{sig-alternate-05-2015}
\usepackage{algorithm}
\usepackage{algorithmic}
\usepackage{listings}
\usepackage{xcolor}
\usepackage{multirow}
\usepackage{threeparttable}
\usepackage{bm}
\usepackage{graphicx}
\usepackage{subfigure}
\usepackage{url}
\usepackage{textcomp}
\usepackage{hhline}
\newcommand{\tabincell}[2]{\begin{tabular}{@{}#1@{}}#2\end{tabular}}

\begin{document}

\title{Target Directed Event Sequence Generation \\for Android Applications}

\author{
	Jiwei Yan$^{1,3}$, Tianyong Wu$^{1,3}$, Jun Yan$^{2,3}$, Jian Zhang$^{1,3}$\\
	\normalsize$^1$State Key Laboratory of Computer Science, Institute of Software, Chinese Academy of Sciences\\
	\normalsize$^2$Technology Center of Software Engineering, Institute of Software, Chinese Academy of Sciences\\
	\normalsize$^3$University of Chinese Academy of Sciences\\
	\normalsize Email:\{yanjw, wuty, yanjun, zj\}@ios.ac.cn
}

\maketitle
\begin{abstract}
Testing is a commonly used approach to ensure the quality of software, of which model-based testing is a hot topic to test GUI programs such as Android applications (apps). Existing approaches mainly either dynamically construct a model that only contains the GUI information, or build a model in the view of code that may fail to describe the changes of GUI widgets during runtime. Besides, most of these models do not support back stack that is a particular mechanism of Android. Therefore, this paper proposes a model LATTE that is constructed dynamically with consideration of the view information in the widgets as well as the back stack, to describe the transition between GUI widgets. We also propose a label set to link the elements of the LATTE model to program snippets. The user can define a subset of the label set as a target for the testing requirements that need to cover some specific parts of the code. To avoid the state explosion problem during model construction, we introduce a definition ``state similarity'' to balance the model accuracy and analysis cost. Based on this model, a target directed test generation method is presented to generate event sequences to effectively cover the target. The experiments on several real-world apps indicate that the generated test cases based on LATTE can reach a high coverage, and with the model we can generate the event sequences to cover a given target with short event sequences.

\end{abstract}

%
%

\ccsdesc[500]{Software and its engineering~Software testing and debugging}
\printccsdesc
\keywords{ Android GUI Model; Targeted Test Generation; Back Stack; Dynamic Modeling}
\newpage
\section{Introduction}
With the success of smart mobile device market, mobile application market ushered a high speed developing period, especially Android application market. Android apps, like other software, need to be adequately tested to eliminate the potential bugs and improve the quality. In the area of testing, an essential step is test case generation, which focuses on how to automatically generate the test suite with high code coverage and strong fault detection ability.

Android apps are event-driven GUI programs. An Android app can be regarded as a collection of widgets, each of which is defined in an \texttt{Activity} class that is provided by Android system to interact with the user. The user operations on the components (corresponding to \texttt{View} class in Android) in the screen trigger the corresponding events to drive the app execute the corresponding code and transfer from one widget to another. Thus, for an Android app under test (AUT), a test input is a sequence of events associated with its widgets.

There are a number of test generation techniques for Android apps, of which model-based testing is an attractive approach for tackling this problem. The general steps of model-based test generation are shown as follows: (1) design a formal and abstract model that can briefly describe the software behavior; (2) translate the software to the model; (3) generate the test cases based on the model. The key point in the model-based testing is how to design and construct a proper model that can accurately and comprehensively describe the behavior of the AUT. In the area of Android apps, the model is often used to describe the Activity transitions of the AUT.

In recent years, several model construction approaches have been proposed for test generation of Android apps, which can be categorized into two kinds, static construction and dynamic one. The former one leverages the static analysis techniques on the code of the AUT to extract the GUI components in each Activity of the AUT and the transitions between Activities. However, this kind of approaches may fail to describe the changes in the screen of one Activity during runtime, e.g., some views are instantiated under the conditions that should be determined dynamically. The latter one regards the AUT as a black-box and makes use of dynamic analysis techniques to ripper the GUI information and trace the transitions between Activities when the AUT is running. As a result, the model constructed by these approaches does not contain code information. In addition, we find that the models proposed in the existing works are not intricate enough to describe the AUT behavior. The existing works often omit several particular mechanisms of Android, such as the back stack with complex launch mode that has a big influence both on model states and transitions.

In this paper, we refine the existing models via comprehensively considering the GUI information and Android mechanisms, and propose a new model called LATTE to describe different states in Activities and the transitions between these states during the execution of an Android app. Most of the existing models distinguish different states according to only the basic information (like the view type and position) of views in the widgets. However, based on our observation, the program behaviour of the same widget will be different when the status of its views and the back stack is different. Therefore, our LATTE model defines a more intricate state that contains the view information as well as the status of views and the back stack to address this issue. On the other hand, too fine-grained state of model may lead to the state explosion problem. We introduce ``state similarity'' to merge similar states to avoid exploring too many states in model construction procedure.

Besides the GUI information, we also link the transitions in the model to the code that are executed when the corresponding GUI events are triggered. We first define a label set and map the code snippets to labels and then mark the transitions of LATTE model with them. We define several labels that correspond to user concern in the model as a target and find a set of pathes based on LATTE to cover it. This model can be better used to generate test cases to cover specific code snippets that the user is concerned about.

We adopt the dynamic construction technique to build our LATTE model from an AUT. We first insert necessary probes into the AUT to record the runtime information related to the back stack and labels. The instrumented AUT is driven to run on a testing framework to explore the GUI components and construct the model on-the-fly. At last, we traverse the model to generate feasible test cases to cover the user given target. We implemented the proposed techniques into a model based testing tool called \emph{AppTag} and compared it with two state-of-art tools Monkey~\cite{Monkey} and Dynodroid~\cite{DBLP:conf/sigsoft/MachiryTN13} (these tools outperform other existing test generation tools~\cite{DBLP:conf/kbse/ChoudharyGO15}).

The main contributions of this work are summarized as follows.
\begin{itemize}
  \item Propose the LATTE model to describe the GUI characteristics in detail for generating test cases.
  \item Provide a dynamic construction approach for LATTE model.
  \item Propose an approach to generate test sequences to cover the user given target.
  \item Implement a model based test generation tool \emph{AppTag} and evaluate it on real-world instances.
\end{itemize}

The remainder of this paper is organized as follows. In Section \ref{Background}, we discuss some necessary background knowledge about Android Activity and event sequence generation. The LATTE model we proposed will be described in Section \ref{model}. In the following section, we will introduce our model construction and target directed test generation approaches in detail. Then we evaluate our approach via the experiments on real-world Android apps in Section \ref{Evaluation}. Section \ref{related} surveys the related work and Section \ref{conclusion} gives the conclusion and discusses the future work.

\section{Background}\label{Background}
The execution of an Android app is composed of a sequence of Activities. Accordingly, a test case of an app is also a series of operations on Activities. In this section, we will present some background knowledge on Activity, and the techniques for event sequence generation, including the GUI ripping technique to extract the views in the widget, and the instrumentation technique for monitoring the operations.

\subsection{Android Activity}
 When an Activity is launched, it will display a widget that is made up with a series of UI views such as buttons and textviews that facilitate the user interaction.  The screen can only be occupied by one Activity, therefore, Android system introduces a \emph{back stack} \cite{backStack} to store all the launched Activities. If a new Activity is activated, the former Activity should be destroyed or pushed into the back stack. In the following part of this subsection, we will briefly introduce the UI views and the back stack.

\subsubsection{UI Views}\label{UI Views}
Android provides various kinds of built-in UI views for user interaction that are all extended from the class \texttt{View} and correspond to different UI events. Besides, Android also allows users to customize the UI views by extending the view and overriding the standard methods. The customized view is usually extended from slight modification or composition of basic views.

The \texttt{View}  class represents the basic building block for user interface components. Except for the type of the component and position displayed on the widget, some views also have several extra attributes during runtime, like enabled, focused or checked. We define these extra attributes as the \emph{status} of the view. In practice, the different statuses of a view may represent different program execution state and may lead to different program behavior (see Section \ref{sec:case_study}). For example, a checkbox which is used for changing the setting of an app with checked and unchecked statuses are totally different. Therefore, in this paper, we consider two views are the same only if all the attributes including the statuses are the same.

A view has several UI events corresponding to different user operations. For example, the possible events for the \texttt{Button} view are \texttt{Click}, \texttt{LongClick} and \texttt{Press}. In addition, some events should be triggered by a combination of several user operations, for instance, the typing event on \texttt{EditText} view needs a text clearing operation followed by a text typing operation. Table \ref{UI types and UI events} shows the commonly used events of UI views according to Android references \cite{View}, where \texttt{Scroll} indicates the scroll operation of \texttt{ListView}, and \texttt{setValue} denotes the operation of setting value for \texttt{ProgressBar}.

\begin{table}[!htbp]
\caption{ UI views and events}\label{UI types and UI events}
\centering
\scriptsize
\begin{tabular}{l|l|l|l|l|l}
\hline
 Type        & \texttt{Click}    & \tabincell{c}{\texttt{LongClick} \\ \texttt{Press}}  & \texttt{Scroll}  & \tabincell{c}{\texttt{ClearText} \\ \texttt{TypeText} }  & \texttt{setValue}     \\
 \hline
 \hline
 \texttt{Button}      &$\surd$   &$\surd$                                 &         &                                                     &              \\
 \hline
 \texttt{RadioButton} &$\surd$   &                                        &         &                                                     &              \\
 \hline
 \texttt{CheckBox }   &$\surd$   &                                        &         &                                                     &              \\
 \hline
 \texttt{ImageView }  &$\surd$   &$\surd$                                 &         &                                                     &              \\
\hline
 \texttt{TextView }   &$\surd$   &$\surd$                                 &         &                                                     &              \\
\hline
 \texttt{EditText}    &$\surd$   &                                        &         & $\surd$                                             &              \\
\hline
 \texttt{ListView}    &$\surd$   &                                        & $\surd$ &                                                     &             \\
\hline
 \texttt{ProgressBar}  &          &                                        &         &                                                     & $\surd$     \\
\hline
\end{tabular}
\end{table}

Besides, there are some global events that can be executed at any program state and do not attach to UI views, such as Rotate, click on the Home, Back and Power key. We consider all these events in analyzing the UI views of an Activity.

\subsubsection{Back Stack} \label{Back Stack}
During the execution of the app, the user interacts with a collection of Activities that are stored in the back stack. When the current Activity starts another one, the new Activity will be pushed on the top of the stack and takes the focus. There is a special feature of Android system that when user presses the hardware back key, the current Activity will be popped and destroyed, and the previous Activity resumes. Activities in the stack can only be rearranged by push and pop operations. Note that an Activity with different back stack may lead to different program behavior (see Section \ref{sec:case_study}).

Android system defines the \emph{Launch Mode} \cite{launchMode} of an Activity to determine the evolution of the back stack when the Activity is launched. There are four launch modes, including Standard, SingleTop, SingleTask and SingleInstance. The launch modes of the Activities are declared in the \emph{Manifest} file of the app or profiled using intent flags in the code. The diversity of launch modes makes the evolution of the back stack complex so that we need to take effort in correctly modeling the back stack. The details of these launch modes will be discussed in Section \ref{launchmode}.

\subsubsection{An Example of Views and Back Stack}\label{tomdroid example}
\begin{figure}[!bp]
\centering
  \begin{tabular}{ccc}
     \includegraphics[width=0.14\textwidth]{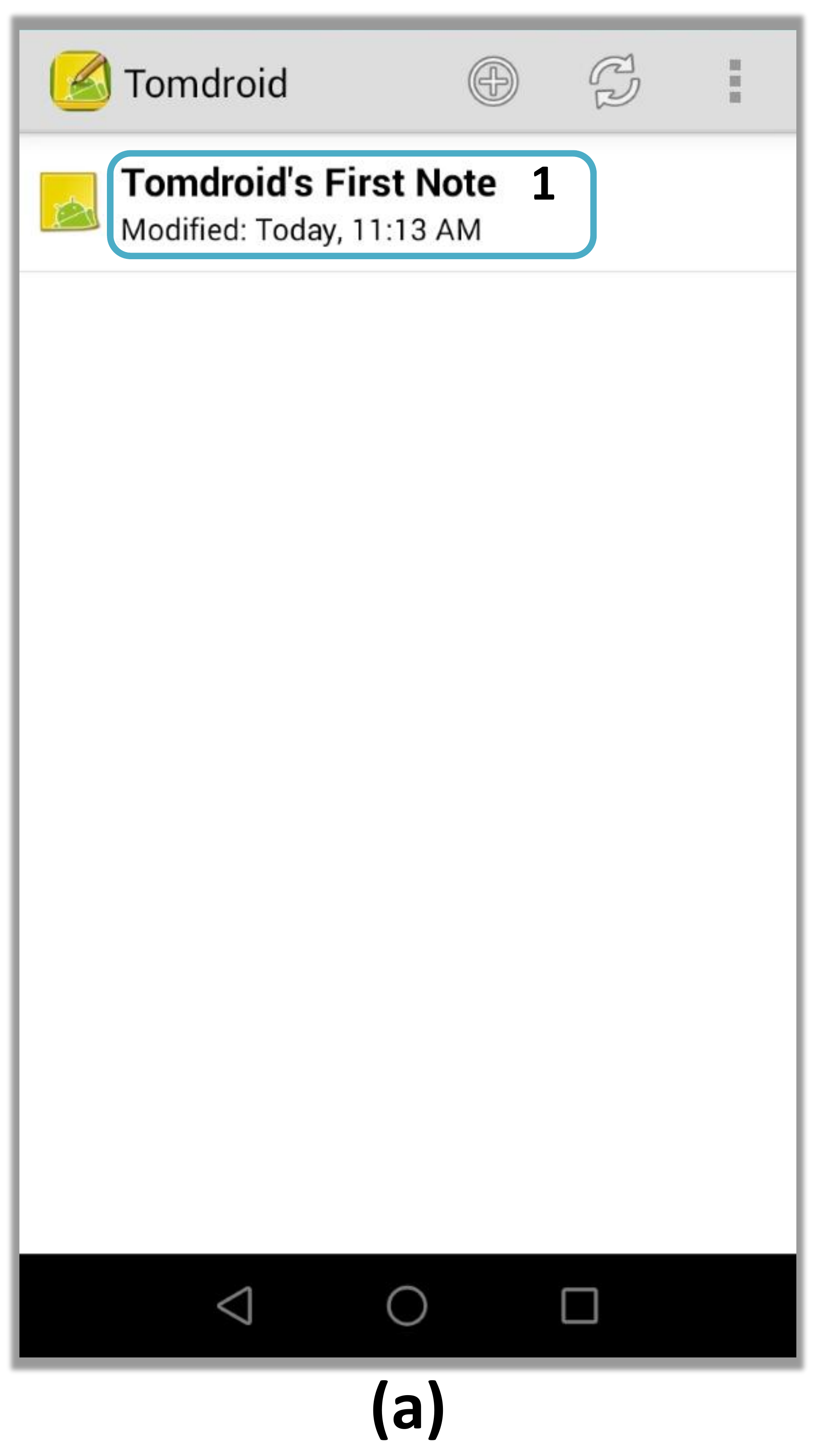}&\includegraphics[width=0.14\textwidth]{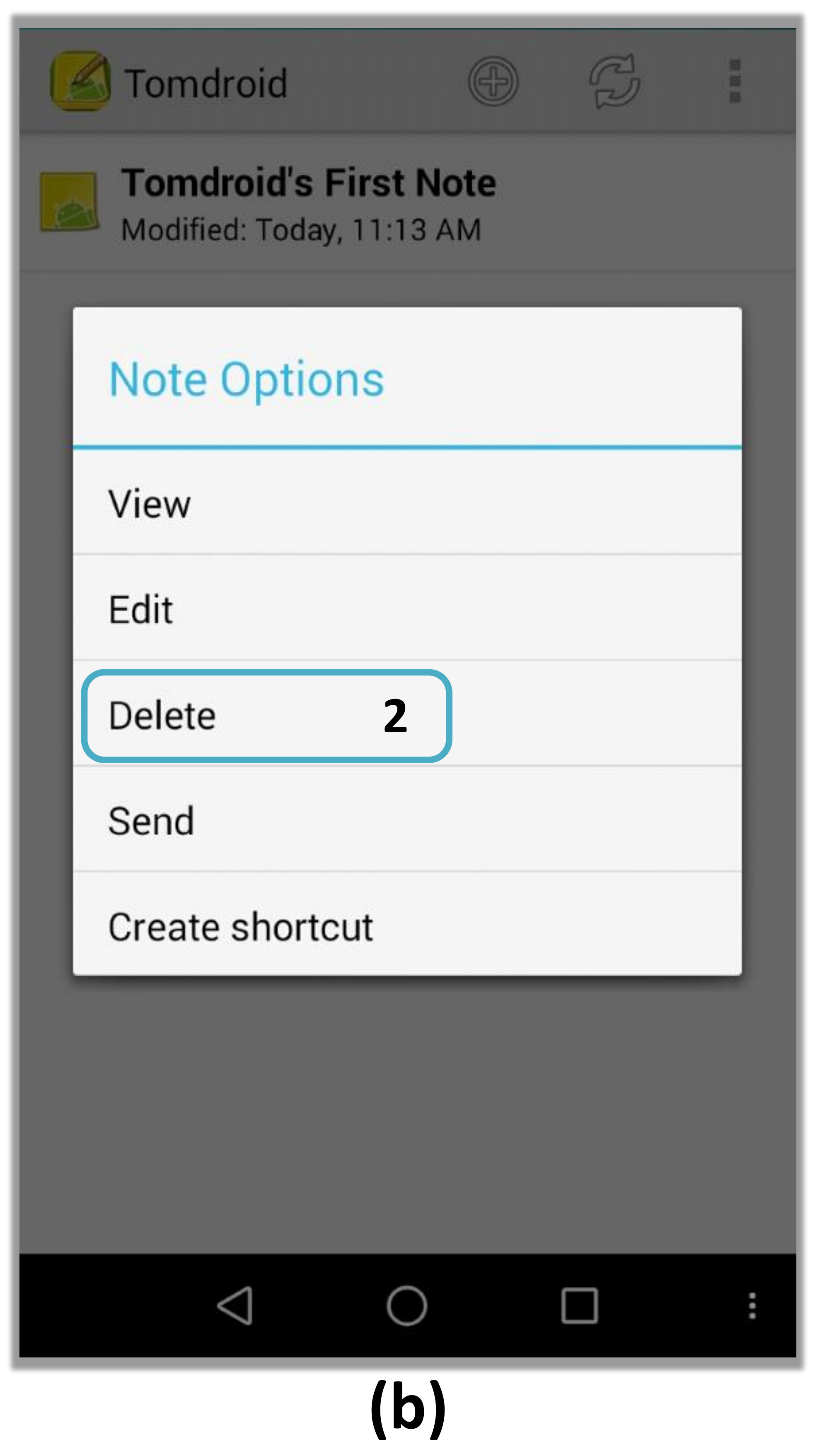} &\includegraphics[width=0.14\textwidth]{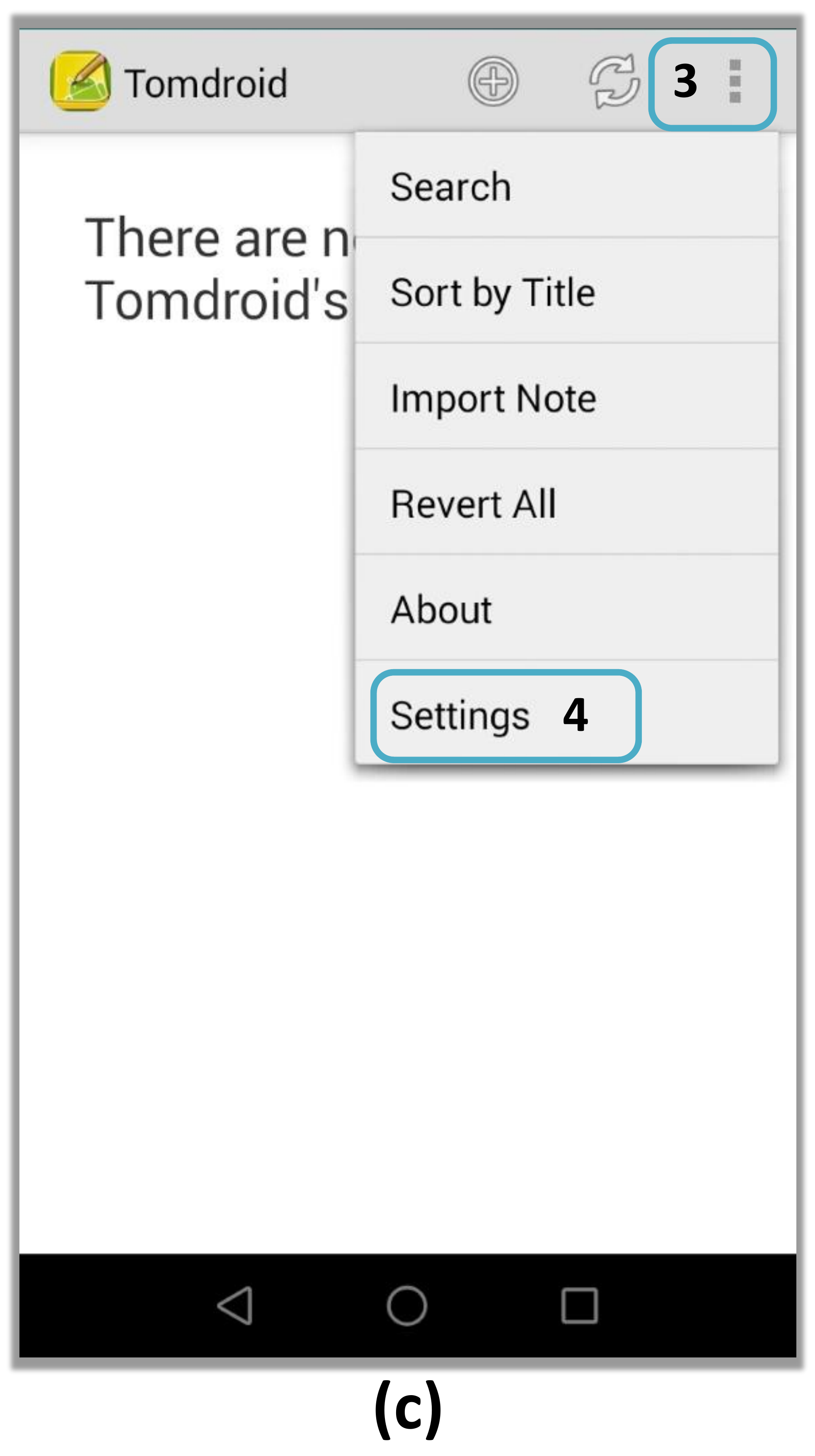} \\
     \includegraphics[width=0.14\textwidth]{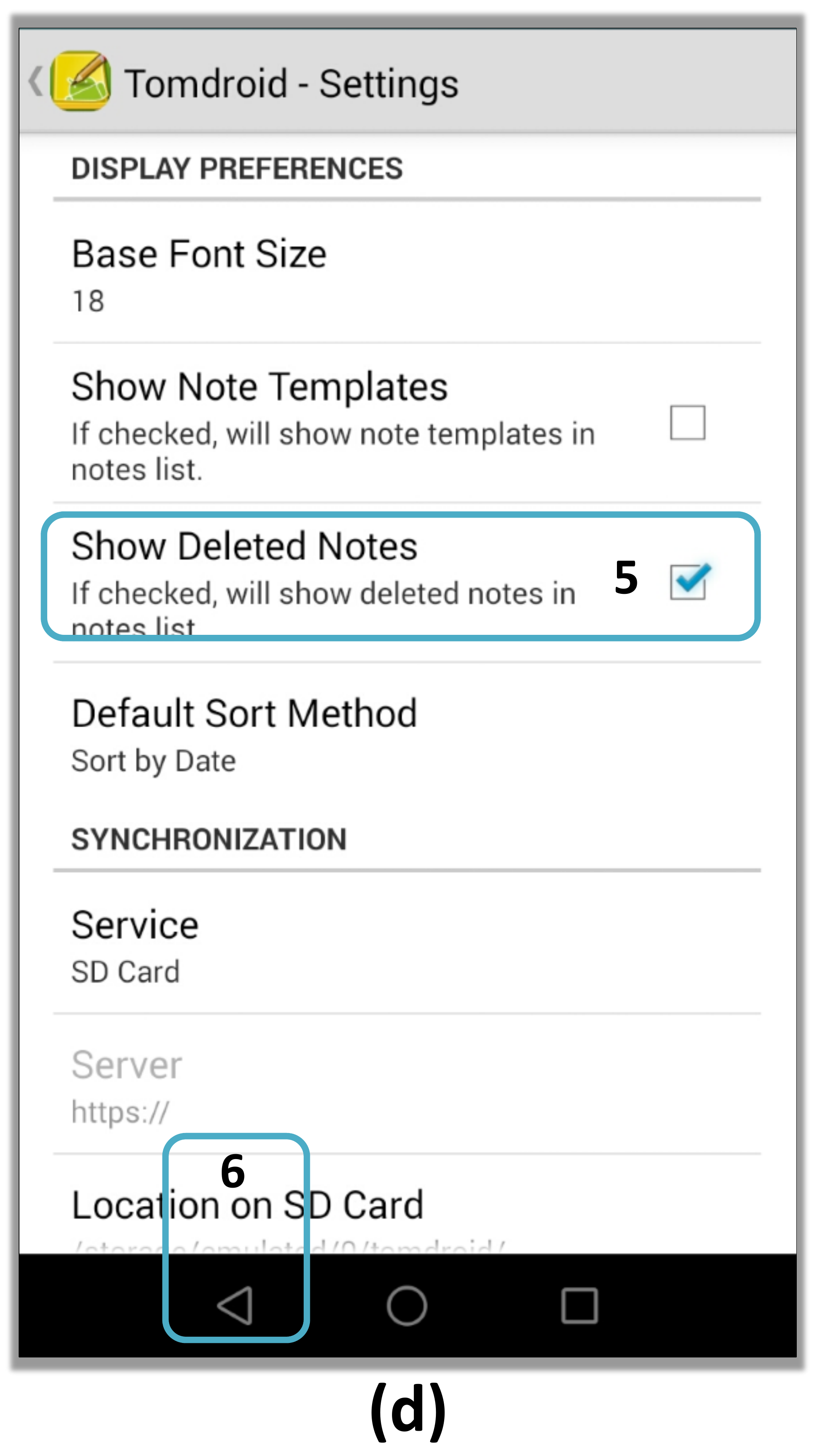}&\includegraphics[width=0.14\textwidth]{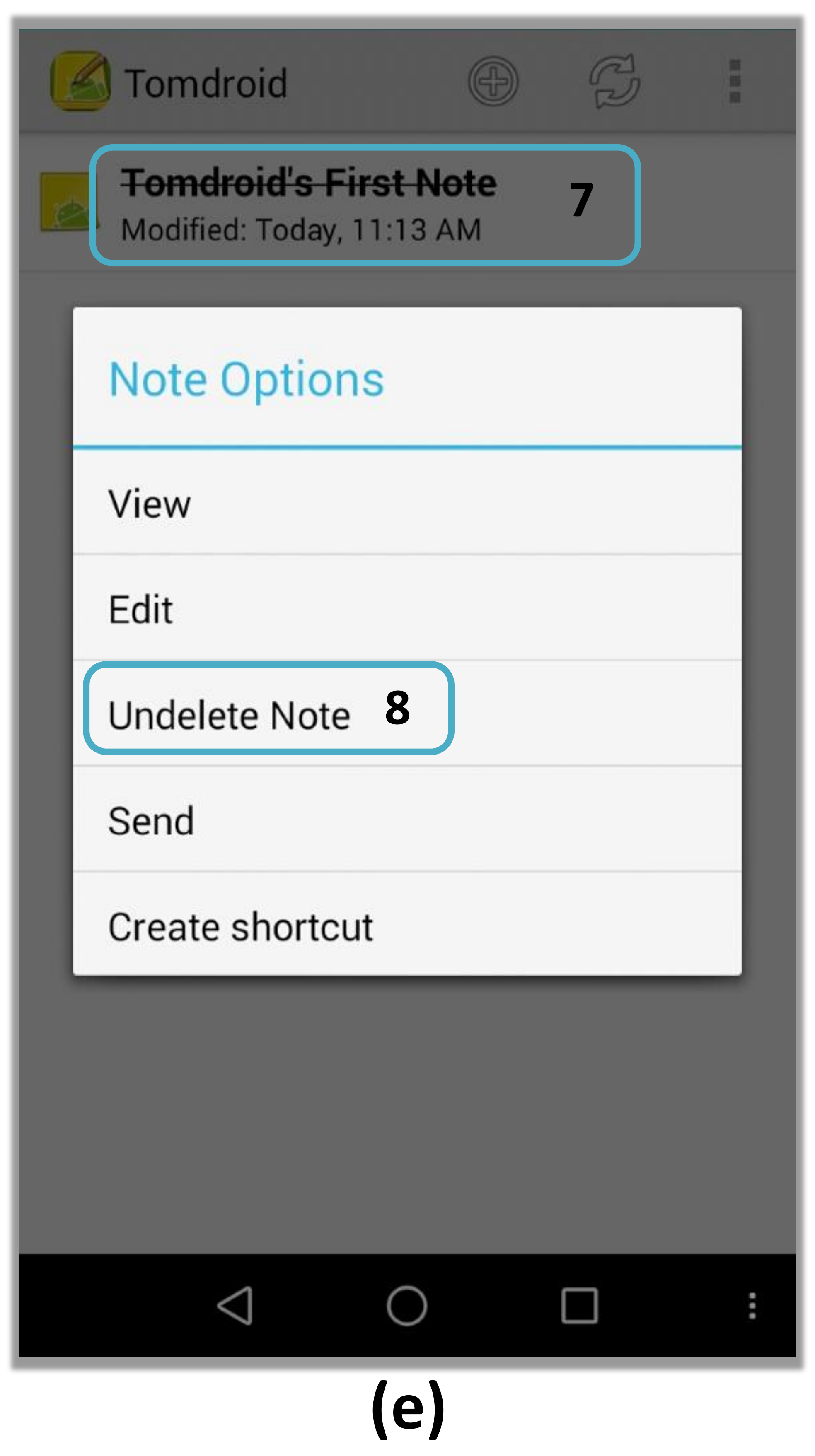} &\includegraphics[width=0.14\textwidth]{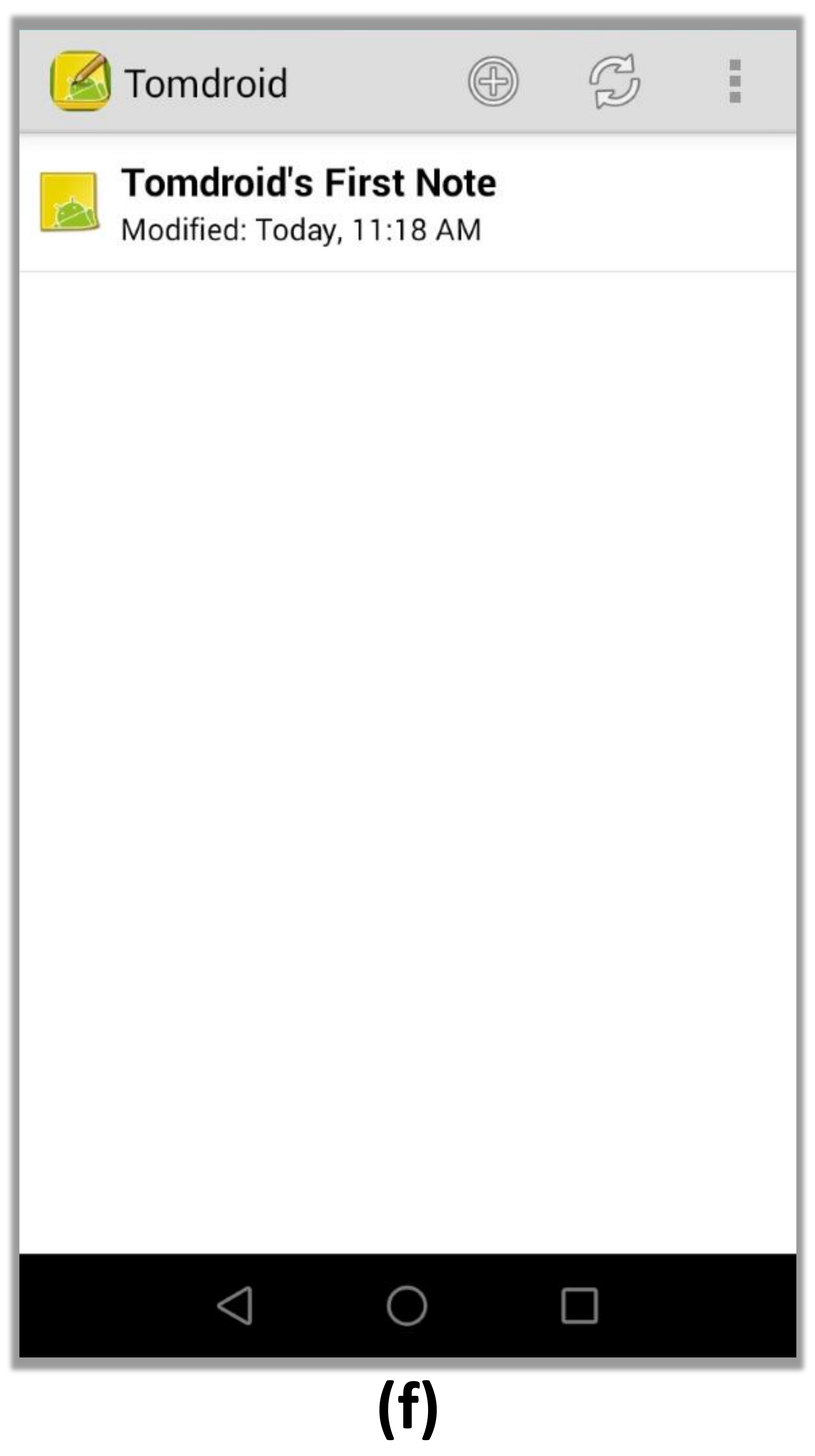} \\
   \end{tabular}
 \caption{TomDroid Application}\label{TomDroid Application}
\end{figure}
\emph{TomDroid} is a popular note-taking application in Android platform. Fig. \ref{TomDroid Application} shows six snapshots (a-f) of two Activities \texttt{TomDroidActivity} and \texttt{PreferencesActivity} in this app, where the eight numbers (1-8) denote a sequence of events that delete a note (1-2), adjust the settings to show all the deleted notes (3-6) and recover the deleted note (7-8). There are several types of views in the widget, including Button, TextView, ImageView, CheckBox, where we can click the Button to jump into another Activity or click the checkbox to change its status. These views correspond to different events described in Table \ref{UI types and UI events}. Besides, in the bottom of the widget, there are three keys that are provided by the Android system. The ``$\lhd$'' key is the hardware back key that will by default drive the app return to the previous Activity or destroy the app when the key is clicked.

\begin{figure}[!htbp]
\centering
\includegraphics[width=0.45\textwidth]{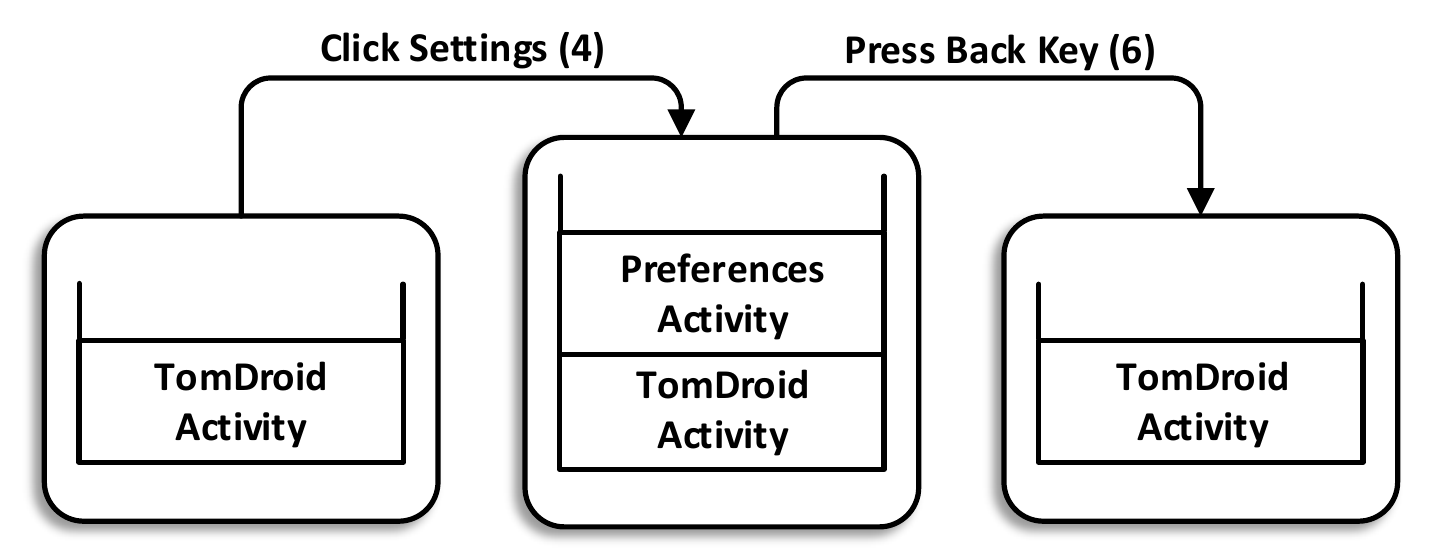}
 \caption{Change of Back Stack}\label{stack graph}
\end{figure}

When the app starts, it will directly jump to the Activity \texttt{TomDroidActivity} (snapshot a) and push this Activity into back stack. After clicking the last menu item ``Settings'' (event 4), the app will start a new instance of Activity \texttt{PreferencesActivity} (snapshot d) and push it into back stack. A click operation can be executed on checkbox (event 5) to change its status, while it will not affect the back stack. After that, if we press the back key (event 6), current Activity \texttt{PreferencesActivity} will be destroyed and popped from the back stack, and previous Activity \texttt{TomDroidActivity} will be resumed. Figure \ref{stack graph} visualizes the evolution of the back stack with the above procedure.

\subsection{Event Sequence Generation}
For the TomDroid example in Section \ref{tomdroid example}, the event sequence from 1 to 8 forms a test case to check the note recovery functionality. Our goal is to generate a set of event sequences that cover such specific code that corresponds to the user concerned functionalities. A simple way like Monkey just randomly sends events to drive AUT to run until the functionality is triggered. Dynodroid improves this idea with some heuristics. However, this type of random approaches are not model based, i.e., they do not construct a global model to record the differences in the status of a same Activity when it is traversed multiple times. As a result, the randomly generated test sequence usually contains a large number of events that burden the workload for debugging.

To generate the compact event sequences, we should construct a model that represents the transitions along with events between the Activities during exploring the GUI to guide the future traversal. Furthermore, this model can carry some deep connections between the status of GUI views and the program logic, such as the back stack, to make it more accurate.

The information of this model can be obtained via the combination of existing dynamic program analysis techniques. The UI views can be extracted by the GUI ripping technique. For the information that can not be captured from GUI, we employ the instrumentation technique that inserts the probes into specific points of the program and collect the concerned information during execution. Here we briefly introduce these two techniques, and the model will be discussed in the next section.

\subsubsection{GUI Ripping}
GUI ripping is a commonly used technique to identify the GUI elements in the GUI programs. There are several frameworks that can obtain the information of views for the Android apps in runtime, such as \emph{Robotium} \cite{Robotium} and \emph{Hierarchyviewer} \cite{HierarchyViewer}. In this paper, we leverage \emph{Robotium} for GUI ripping.

\emph{Robotium} is an open source Android test framework that drives the AUT to execute under manually designed test cases. It is based on the test instrumentation mechanism \cite{ActivityInstrumentationTestCase2} provided by Android system, that can monitor the interaction of application and Android system, and control the execution of AUT. Robotium also provides a series of APIs to capture and access the views and send instructions to simulate user events.

\subsubsection{Instrumentation Technique}
Besides the information of GUI views, we also need to obtain some code information to construct our model. Therefore, we make use of the instrumentation technique to monitor the code snippets that we are interested in. This approach modifies application's original code by adding a few statements into it. The statements inserted are called program probes. They are usually added for examining the execution of program statements and runtime change of variables. These probes will work during the dynamic running of instrumented application, and provide runtime information of program statements.

The instrumentation technique is often implemented on the source code, but the source code is not always available especially for apps in Android markets. Hence, we perform instrumentation on the ``smali'' byte-code, a readable format of the byte-code of the app. The smali format provides multiple keywords, e.g., the \texttt{.method} keyword denotes the beginning of a method body and \texttt{invoke} keyword is used to invoke the indicated method. We should put the probes properly according to the grammar and semantic context of original program. In addition, instrumentation on byte code should take good care of registers to avoid conflicts and ensure that the application shall run normally after modification.

\section{Labelled Activity Transition \\ Model}\label{model}
Many model based test generation research works make use of some abstract model to describe the AUT, for example, finite state machine (FSM) is widely used in previous works \cite{DBLP:conf/kbse/AmalfitanoFTCM12, DBLP:conf/oopsla/AzimN13,DBLP:conf/oopsla/ChoiNS13,  DBLP:conf/mobisys/Hao0NHG14,DBLP:conf/fase/YangPX13}. However, these works ignore either the back stack or the statuses of UI views, that leads to an imprecise UI model. On the one hand, a model without back stack has difficulty in generating test cases containing back operations, as the destination of back operation is determined by information of back stack. On the other hand, the change of view status may have influence on the app's running state. Without the consideration of view status, some features of apps will be omitted.

In this section, we will describe a \emph{Labelled Activity Transition graph with sTack and Events} (LATTE) model and give a similarity detection approach for constructing accurate app models.

\subsection{Model Formalization}
As discussed above, an Activity may contain multiple suites of views and events in runtime, that may lead to the loss of some program states when we regard an Activity as a state. Therefore, we propose a LATTE model, whose main idea is to split each Activity into one or more states, depending on the current views and events in the screen and the back stack information. The LATTE model can be formally described as a 5-tuple $\mathcal{M}=\langle S, La,T, s_0, q\rangle$, where
\begin{itemize}
  \item $S$ is a finite set of application's runtime states. An element $s \in S$ is a triple $\langle a, V_s, L_s \rangle$ where $a$ denotes the Activity that $s$ corresponds to, and $V_s$ indicates the set of views that belong to the state, and $L_s$ is a list of Activities in the back stack. (We have not addressed the events here, the reason is that for a concrete state split from an Activity, the events are bound to the set of views.)
  \item $La$ is the set of labels.

  \item $T$ denotes the set of transitions. Each element $t \in T$ is a 4-tuple $\langle src, e, la, des \rangle$ representing the transition from the source state $src$ to the destination state $des$ caused by event $e$ bound to the state $src$, and the label set $la \subset La$ denotes the labels assigned to this transition.
  \item $s_0 \in S$ is the entry state that represents the initial state of the app.

  \item $q$ is a terminal state denoting that the app quits or jumps to another app.

\end{itemize}

Note that the denotations with subscript $s$, including the view set $V_s$ and the back stack $L_s$, describe the attributes that are affiliated to the state $s$. We should have regarded the states that have the same Activity and a minor change in the view set or the back stack as different states. However, in the real-world Android apps, this way may cause a state explosion problem during the model construction. In order to balance the performance of model accuracy and the analysis cost of AUT, we propose a ``state similarity'' measurement to merge the similar states in the next subsection.

\subsection{State Similarity}
According to the definition of the LATTE model, a state is a triple $\langle a, V_s, L_s \rangle$. We measure the similarity of two states derived from the same Activity according to the views and back stack, of which views can be characterized with two attributes, the list of GUI components and view status of each view (see Section \ref{UI Views}).

Let $s_1$ and $s_2$ be two states, and two functions $Sim_{V}$ and $Sim_{L}$ calculate the similarity of the view and back stack. We have the following formula to compute the state similarity of two states $s_1$ and $s_2$.
$$Sim(s_1, s_2) = \omega \cdot Sim_{V}(s_1, s_2)+ (1.0-\omega) \cdot Sim_{L}(s_1, s_2)$$
Where the function $Sim_{V}$ returns the percentage of same views in all the views of the two states, i.e., $Sim_{V}(s_1, s_2) = |V_1 \bigcap V_2| / |V_1 \bigcup V_2|$ where $V_1$ and $V_2$ denote the set of views of $s_1$ and $s_2$ respectively. The function $ Sim_{L}$ returns 1 if the back stacks of the two states are the same, otherwise 0. We define the configurable parameter $\omega$ ($0 \leq \omega \leq 1.0$) to represent the weights of these two functions.

Two states with high similarity can almost be regarded as the same. Therefore, in our approach, we introduce the similarity threshold $S_T$ to avoid too many states split from an Activity. When building the LATTE model for an app, if the similarity of the newly explored state and an existing state in the model is higher than $S_T$, we will not introduce a new state in the model. Instead, We will merge the new state to the existing state that is most similar to it.

\subsection{Label and Target}
We introduce a label set $La$ to embed the code information into our model. In general, each element in $La$ corresponds to a part of specific code. For example, we can set each label to represent a distinct method of the AUT, or all the methods in the same class. The mapping rule for the labels and the code snippets are designed according to the actual testing requirements. A fine-grained rule can make the model contain more accurate code information while it increases the model scale and the cost of the model construction. With the mapping rule, the labelling procedure is implemented via code instrumentation in our approach.

The motivation of this work is mainly inspired from the observation that the testers are often concerned about some specific parts of code in the AUT. For example, when the testers want to test a functionality of the AUT, they only focus on several methods related to this functionality. We call the set of these specific code the ``target''. Formally speaking, in this work, we define the target as a set of labels in the label set $La$. For example, if we have a label set $La = \{la_1, la_2, la_3\}$, we can set the target as $Ta = \{la_1, la_3\}$ representing that the test cases should cover all the transitions labeled with $la_1$ and $la_3$.

In the testing procedure, many kinds of subjects can be regarded as targets. In this paper, we just consider two types of targets, including covering a set of specific user-developed methods and a set of system APIs related to resource and privacy. The former one can make sense in the above situation when the testers want to test a specific functionality implemented by developers. The latter one focuses on two kinds of important system APIs, where the misuse of the APIs for the resources (Camera, Media Player and Sensors, etc.) will lead to performance decrease \cite{TSE16relda}, and the privacy related APIs are essential to the application security \cite{Arzt14PLDI}. By labelling these targets, we can guide our approach to generate possible test cases that can reach these labels or the combinations of them. Execution of these test cases can trigger the bugs caused by specific parts of the program, or reveal some insecure operations of the app.

\subsection{LATTE Model of TomDroid}

\begin{figure}[!htbp]
\centering
\includegraphics[width=0.43\textwidth]{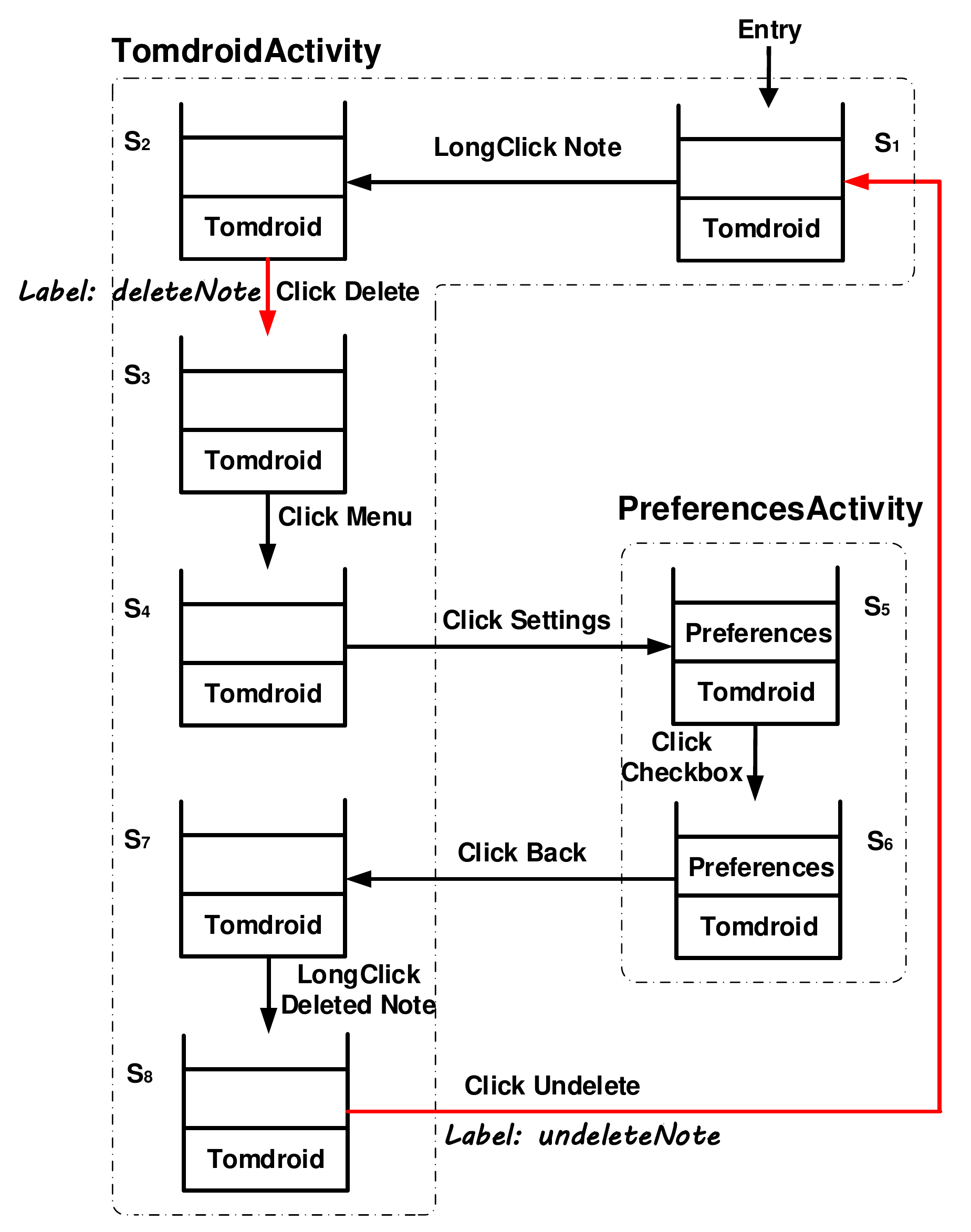}
\caption{LATTE Model of TomDroid}\label{LATTE Model of Tom}
\end{figure}

Fig. \ref{LATTE Model of Tom} shows parts of the LATTE model of TomDroid in Fig. \ref{TomDroid Application}. Each entity indicates a state that is marked by its state id as well as its back stack. For simplicity, we do not give the view information of each state in the figure. The states in the same dashed box correspond to the same Activity. Each edge from one state to another is a transition corresponding to an event in Fig. \ref{TomDroid Application} and the red solid edge illustrates the label set of the transition is not empty. Here we deem label set $La=\{la_1, la_2\}$ where $la_1$ and $la_2$ represent the invocation of the methods \texttt{deleteNote} and \texttt{undeleteNote} respectively.

Note that in state $s_5$ (snapshot d), the status of checkbox ``Show Deleted Notes'' of Activity \texttt{PreferencesActivity} will affect the layout of Activity \texttt{TomDroidActivity} that if it is not checked, the deleted notes will not appear and can not be recovered. Therefore, it is necessary to split the Activity \texttt{PreferencesActivity} into two states ($s_5$ and $s_6$).

\section{Proposed Approach}\label{approach}
Given an AUT and a target, our goal is to generate a small set of executable event sequences for covering the target. In this section, we will first give an overview of the approach, and then discuss the construction of the LATTE model and propose an adaptive test generation approach.

\subsection{Overview}
\begin{figure}[!htbp]
\centering
\includegraphics[width=0.43\textwidth]{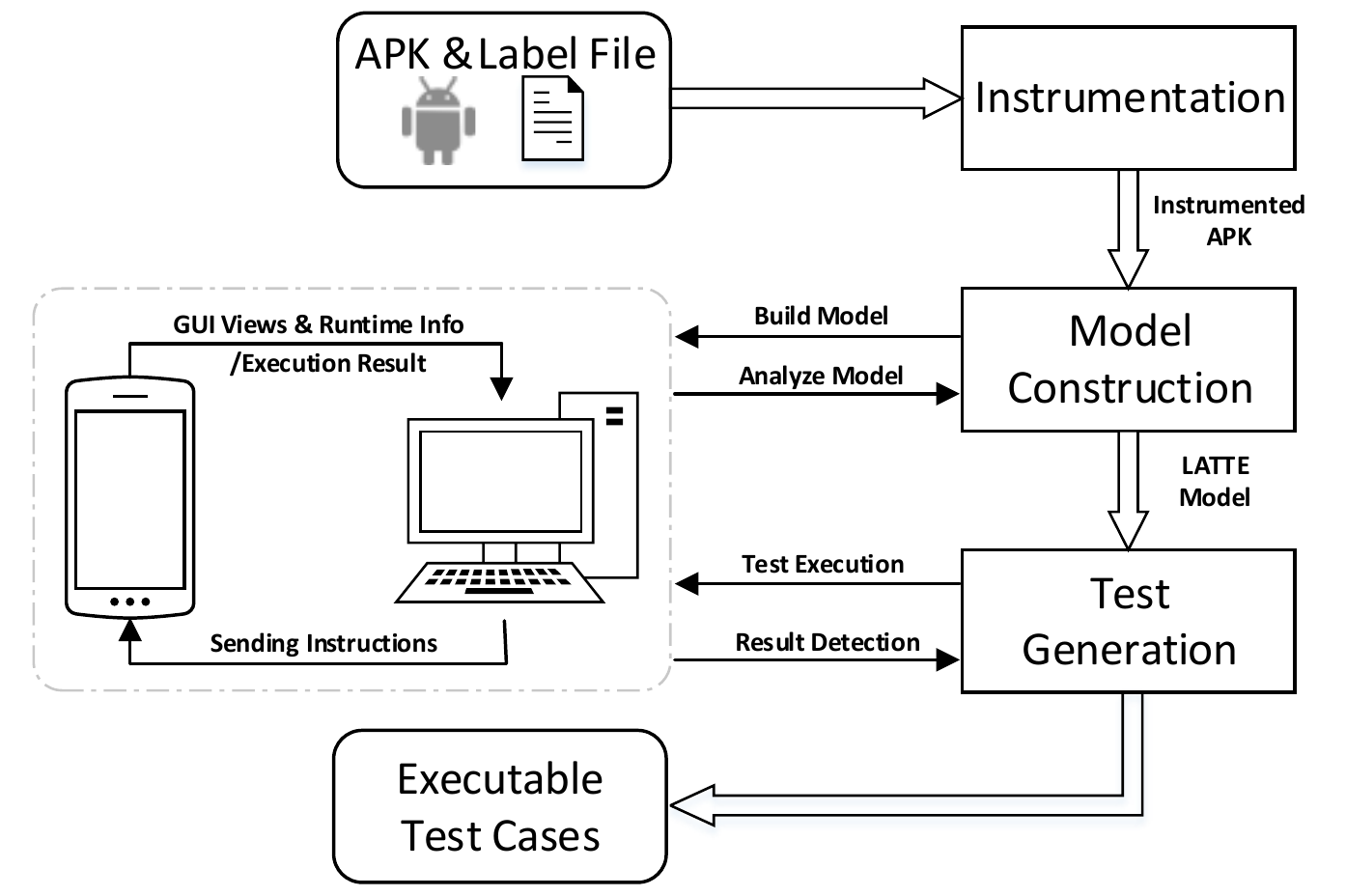}
\caption{Approach Overview}\label{Approach Overview}
\end{figure}

Fig. \ref{Approach Overview} shows the overview of our approach, which is composed by three parts: Instrumentation, Model Construction and Adaptive Test Generation.

Our approach takes an Android apk file and a label set as the input. Firstly, we instrument probes into the AUT to monitor the runtime information for target labelling and model generation. Secondly, we drive the instrumented AUT running on the Android device and exploring the AUT for constructing the LATTE model. Then we generate executable test cases for the user given target based on the model by a modified traversal algorithm with some heuristics. In this step, a feedback approach is used to check the feasibility of test cases by monitoring the execution.

\subsection{Instrumentation}

 We adopt a light-weight instrumentation method on bytecode to insert probes monitoring the execution of code about Activity transition, back stack and labels. First of all, we scan the files decompiled from apk file to locate the statements related to targets and the creation and destroy operations on Activity. Next, we add probes following the statements for logging their run-time information.

\subsection{Model Construction}

After we have got an instrumented app, we make use of the existing Android testing framework and develop a script to drive the app running on a device and construct the LATTE model on-the-fly according to the log information. The script performs an iterative operation of app exploration and model construction. During each iteration, our script obtains the views and status information of the Activity at the top of the back stack, collects the candidate of possible events according to the views (refer to Table \ref{UI types and UI events}), and selects the next event and send to the testing framework. The iteration will terminate when all events have been triggered. In the rest of this subsection, we will describe our algorithm, and show how we handle the back stack and launch mode.

\subsubsection{Traversal Algorithm}\label{Traversal Algorithm}
Algorithm \ref{LATTE Model Construction} shows the details of the model construction that is based on the BFS traversal algorithm.

\begin{algorithm}[!htbp]
\caption{LATTE Model Construction}\label{LATTE Model Construction}
\begin{algorithmic}[1]
    \REQUIRE Instrumented Apk File, Label Set 
    \ENSURE LATTE model $\mathcal{M}$
    \STATE \textbf{var:} Queue<State> $q$
    \STATE initialize $q$ with the entry state $s_0$
    \WHILE{$q$ is not empty}
        \STATE get the first state $s_h$ from $q$
        \STATE drive the app to the state $s_h$
        \STATE perform an unvisited event $e$ of $s_h$
        \STATE obtain the label set $la$
        \STATE create a new state $s_n$ with logging information
        \STATE create a new transition from $s_h$ to $s_n$ with $e$ and $la$
        \STATE find state $s_m$ that has maximum similarity with $s_n$
        \IF{$Sim(s_m, s_n) > S_T$}
            \STATE merge $s_n$ to $s_m$
        \ELSE
            \STATE append $s_n$ to queue $q$
        \ENDIF
        \IF{all events of $s_h$ have been visited}
            \STATE remove $s_h$ from $q$
        \ENDIF
    \ENDWHILE
\end{algorithmic}
\end{algorithm}

The algorithm maintains the model $\mathcal{M}$ of explored part of AUT and a queue $q$ storing unvisited states. The exploration starts from the entry state $s_0$ and ends when all states have been visited. In each iteration, we first get the front state $s_h$ from the queue $q$ and drive the app to the state $s_h$ according to the event sequence from the entry state to the state $s_h$ that we record when $s_h$ is detected. Next, we select an unvisited event $e$ of $s_h$ and execute it with this event to a new state $s_n$. Then we collect the runtime information of $s_n$ and create a transition from $s_h$ to $s_n$. The event $e$ and the labels $la$ related to target are also assigned to this transition. Then we will calculate the similarity values of $s_h$ and existing states in the model $\mathcal{M}$. If the maximum similarity value exceeds the threshold $S_T$ , we will merge the new state to the existing state that is most similar to it, otherwise, we will append $s_n$ to queue $q$. If all events of $s_h$ have been visited, we will remove $s_h$ from $q$.

\subsubsection{Launch Mode Analysis}\label{launchmode}
In Algorithm \ref{LATTE Model Construction}, we need to create a new state $s_n$ with logging information in Line 8. According to the definition of LATTE, a state is determined by its Activity, views and the back stack, of which the back stack can not be obtained directly from existing Android testing framework. In our approach, we maintain the back stack information for each state by monitoring the APIs \texttt{onStartActivity()} and \texttt{finish()} which will be invoked when a new Activity is started and destroyed. The default operation of \texttt{onStartActivity()} is pushing new Activity instance into the back stack, but it may be complex when the launch mode is considered. We have addressed the influence of launch modes on back stack in section \ref{Back Stack}. Our work is a test generation for a single app, thus we focus on the \texttt{Standard}, \texttt{SingleTop} and \texttt{SingleTask} modes, for the rest one \texttt{SingleInstance} mode is often used in cross-app transitions.

\begin{table}[!htbp]
\caption{Rules for Back Stack Changing}\label{Back Stack Analysing}
\centering
\begin{tabular}{l|l|l|l}
\hline
Launch Mode &Event & Stack Bef. &Stack Aft.\\
\hline
\hline
\texttt{Standard}       &Open a  &($\dots$,a)             &($\dots$,a,a)  \\
\texttt{Standard}       &Open b  &($\dots$,a)             &($\dots$,a,b)  \\
\texttt{SingleTop}      &Open a  &($\dots$,a)             &($\dots$,a)    \\
\texttt{SingleTop}      &Open a  &($\dots$,a,b)           &($\dots$,a,b,a)  \\
\texttt{SingleTask}     &Open a  &($\dots$,a,b)           &($\dots$,a)  \\
\texttt{SingleTask}     &Open c  &($\dots$,a,b)           &($\dots$,a,b,c)  \\
\hline
\end{tabular}
\end{table}

Here we will discuss the rules for handling launch modes in Table \ref{Back Stack Analysing} where the letters a, b and c denote different instances of Activity. The \emph{Standard} mode simply pushes and pops the new launched Activity without considering the same Activity in the stack, i.e., one Activity can be instantiated multiple times (the first row). The difference between \emph{SingleTop} mode and \emph{Standard} lies in that when the Activity is already on the top of the back stack, the \emph{SingleTop} mode refuses to create a new instance of the Activity (the third row). The \emph{singleTask} mode does not allow multiple instances of one Activity in a task, in other words, if an instance of the Activity already exists, all Activities above it will be popped and current Activity will be stored on the top of the back stack (the fifth row).

During the model construction, we maintain a global stack to simulate the Android back stack of the app. When a new state is created (Line 8 in Algorithm \ref{LATTE Model Construction}), we determine whether a new instance of Activity is started. If it is, we obtain its launch mode and update the global stack according to the rules in Table \ref{Back Stack Analysing}. Finally, we assign the information of the global stack to the back stack of the new state.

\subsection{Adaptive Test Generation}
In this subsection, we will introduce our adaptive target directed test generation procedure in the following Algorithm \ref{Adaptive Test Generating}.
We take the LATTE model $\mathcal{M}$ and a target label set $Ta$ whose elements are labels to be covered as input.
We first construct the set $LT$ of transitions whose label set contains at least one element in $Ta$.
Then we leverage several graph algorithms on the model $\mathcal{M}$ to extract the dependency relationships between states and transitions.
Combining these information and the transition set \emph{LT}, we can obtain some reachability information of states and transitions, like which state can directly or indirectly reach some transitions in \emph{LT}, for assisting further adaptive test generation.
For a transition $l \in LT$, we try to find an event sequence $e_s \notin TES$ that covers it by a modified DFS algorithm with some heuristics based on the above information. For example, the next transition that could lead to cover most labels in target has the highest priority to be selected.
The sequence $e_s$ will be converted into a test script which can be deployed and running on the device.
After the execution, a running result will be returned that determines whether $e_s$ is a feasible path or not.
If the execution of current path failed, we attempt to find another path executable.
For each transition in $LT$, we set a \emph{MAXTRY} limit to restrict the times of attempts.

\begin{algorithm}[ht]
\caption{Adaptive Test Generation}\label{Adaptive Test Generating}
\begin{algorithmic}[1]
    \REQUIRE LATTE model $\mathcal{M}$, Target Label Set $Ta$
    \ENSURE Targeted Event Sequence Set \emph{TES}
    \STATE  \emph{TES}$ \leftarrow \Phi$
    \STATE construct the set $LT$ of transitions labeled with $Ta$
    \FOR{each $\ell\in LT$ }
        \STATE $try \leftarrow 0$
        \WHILE{$try < $ \emph{MAXTRY}}
            \STATE $try \leftarrow try +1$
            \IF{  there exists an event sequence $es \notin$ \emph{TES}  that covers $\ell$}
                \IF{$es$ is executable}
                    \STATE add $es$ to \emph{TES}
                \ENDIF
            \ELSE
                \STATE break
            \ENDIF
        \ENDWHILE
    \ENDFOR
    \end{algorithmic}
\end{algorithm}

\section{Evaluation}\label{Evaluation}
To evaluate the effectiveness of our approach, we implement a tool AppTag on the top of the testing framework Robotium to construct model and generate target directed test sequences for AUT. All of our experiments are done on a Samsung I9300 cellphone, with its 1.4GHz CPU, 1GB RAM, and 16G ROM. The AUT will be reinstalled after each test to ensure the same initial environment for each testing process.

\subsection{Experimental Setup}
To evaluate the effectiveness of our approach, we raise several research questions as follows.
\begin{itemize}
	\item RQ1. Can the GUI exploration approach achieve a high code coverage of apps?
	\item RQ2. How does the similarity impact the model size and coverage?
	\item RQ3. What is the effectiveness of our approach to generate event sequences for covering the given target?
\end{itemize}

To answer these questions, we collect 20 real-world apps as experimental instances. Tool AppTag is applied to construct LATTE model for them and a series of experiments are conducted on them. Table \ref{Experimental Applications} lists the detailed information of these experimental instances, of which the first ten apps are from F-Droid \cite{F-Droid} (with source code) and the rest ones are from the commercial market (without source code). The first column denotes the name of an app. The following four columns show the size (MB) of the app, the numbers of its classes (\#C), methods (\#M), and Activities (\#A). The last three columns give the numbers of Activities whose launch mode is not \texttt{Standard} (\#NS), back stack related API calls (\#B) and widgets that have dynamic status attributes (\#W).


\begin{table}[!ht]
	\scriptsize
	\caption{Experimental Applications}\label{Experimental Applications}
	\centering
	\begin{tabular}{l|c|c|c|c|c|c|c}
		\hline\hline
		\textbf{App } &\textbf{Size}&\textbf{\#C}&\textbf{\#M}&\textbf{\#A}&\textbf{\#NS}&\textbf{\#B}&\textbf{\#W}\\
		\hline
		aGrep			&	0.34	&46		&174	&6	 &2    &21   &3  \\
		aLogcat 		&	0.14	&35		&185  	&2   &1    &15    &3  \\
		BookCl	        &	2.73	&877	&4361	&35  &2    &174   &15  \\
		Budget 		    &	0.19	&63		&272	&8	 &0    &27    &1  \\
		HotDeath 		&	7.93	&28		&355	&3	 &0    &4     &0  \\
		PassWordMP	    &  1.67      &89 	&452	&8   &0   &43   &13	\\
		TippyTipper  	&	0.09    &44	  	&226  	&5   &0    &7     &3  \\
		TomDroid		&	1.08	&154	&834	&8   &0    &54    &8  \\
		WebSearch 		&	1.90	&45		&176	&3   &0    &46    &2  \\
		WhoHasMS 	    &	0.79	&24		&139	&2	 &0    &32    &3  \\
		\hline
		Btime     	    &  14.86     &3752	&25641	&217 &12  &1037 &21	\\
		BubeiListen 	&	3.84	&902	&4637	&91  &8    &491   &0  \\
		Compass 		&	1.38	&29		&316	&2   &0    &69    &0  \\
		Cradio 	        &	1.57	&43		&486	&6   &0    &72    &3  \\
		Flashlight	    &  5.44      &91	&479	&11  &4   &72   &4	\\
		FreshBowser	    &  2.29      &164	&463 	&7   &1   &29   &3	\\
		QiuShiBaiKe    	&  14.31     &2803	&15482	&146 &9   &729  &21	\\
		SaoleiGame    	&  0.34      &23 	&111 	&3   &0   &13   &4	\\
		SgSearch 	    &	8.42	&272	&1083	&66  &8    &305   &26  \\
		Terminal 		&	11.67	&6		&24		&4   &0    &88    &2  \\
		\hline\hline
	\end{tabular}
\end{table}

For RQ1, we measure the effectiveness of GUI exploration approach with the code coverage on the behavior of apps.
The code coverage can be calculated by analyzing the basic information of byte-code and collecting the runtime information of executed code.
We pick two popular automatic testing tools Monkey \cite{Monkey} and Dynodroid \cite{DBLP:conf/sigsoft/MachiryTN13} for comparison, since a recent research~\cite{DBLP:conf/kbse/ChoudharyGO15} shows that Monkey and Dynodroid achieve higher coverage than other existing testing tools for Android apps. The number of generated events for Monkey is 10000 and for Dynodroid is 2000 (same with what Machiry et al. suggested in their work). The similarity threshold $S_T$ of AppTag is set to an experimental value 0.8 (refer to Section \ref{similarity}).

For RQ2, we design experiments to show that how the similarity setting of states influences the size of LATTE model.
In these experiments, we set the value of threshold $S_T$ from 0 to 1.0 and compare the number of transitions in the generated LATTE model and the code coverage by traversing this model under different values of $S_T$. A higher threshold $S_T$ may cause more state splitting and lead to a more accurate model, and further lead to a higher code coverage.

The generated test cases based on LATTE can be manually adjusted for specific testing goals.

For RQ3, we utilize LATTE model for test generation. In practice, the testers often concern some specific part of codes, for example, when the testers want to analyze a functionality of the app, they only focus on the methods related to this functionality. We call the set of these specific code the ``target'', which is a subset of labels in the label set $La$.
We consider two types of targets in this paper, including a set of specific user-developed methods and a set of system APIs related to resource and privacy. The misuse of the latter will cause performance and security problems \cite{DBLP:journals/tse/WuLXGZYZ16,Arzt14PLDI,DBLP:conf/icse/0029BBKTARBOM15}.
Experiments are done between AppTag and Monkey to compare the minimal sequence length they need to cover the given target. To get the minimal sequence length of Monkey, we implement a script to repeatedly run Monkey with the event limits increased by 1000 in each iteration, until the given target is covered.

\subsection{The Code Coverage of GUI Exploration}
In this section, we first generate the test suites for each experimental instance by Monkey, Dynodroid, and our approach respectively, and then calculate the coverage of these three test suites. For the apps with the source code, we calculate the method coverage (\#MC) and line coverage (\#LC) by \emph{EMMA} \cite{Emma}, a code coverage measurement tool for Java programs. For the commercial apps without source code, there is no publicly available code coverage measurement tool. So we make use of the tool InsDal to record the executed code information during runtime and calculate the class coverage (\#CC) and method coverage on byte-code.

\begin{table}[!ht]
	\scriptsize
	\caption{Coverage Comparison on Dalvik Byte-Code}\label{Experimental Applications 1}
	\centering
	\begin{tabular}{l| cc|cc|cc}
		\hline\hline
		\multirow{2}{*}{\textbf{App}} &\multicolumn{2}{c|}{\textbf{AppTag} }&\multicolumn{2}{c|}{\textbf{Monkey}} &\multicolumn{2}{c}{\textbf{Dynodroid}} \\
		&\textbf{CC} & \textbf{MC} & \textbf{CC} & \textbf{MC}  & \textbf{CC} & \textbf{MC} \\
		\hline
		aGrep			&83		&52			&58		&33		&76		&58		\\
		aLogcat 		&74		&67		 	&65 	&58 	&72		&64    \\
		BookCl	        &51		&41			&27		&24		&30 	&26		\\
		Budget 		    &76		&65			&59		&52		&--		&--		\\
		HotDeath 		&86		&79			&68		&54		&85		&72		\\
		PassWordMP      &74		&57			&67 	&52 	&--	    &--	    \\
		TippyTipper  	&93 	&77 		&55 	&58  	&-- 	&-- 	\\
		TomDroid		&60	  	&42 		&38 	&32 	&58 	&40 	\\
		WebSearch 		&69 	&58			&64 	&49	 	&62		&57 	\\
		WhoHasMS	    &91	    &57		    &75		&44		&--		&--		\\
		\hline
		Btime      	    &37		&22			&10  	&6  	&--	    &-- 	\\
		BubeiListen 	&55		&53			&35 	&32 	&19 	&11		\\
		Compass 		&53 	&21 		&55 	&24 	&48 	&14 	\\
		Cradio 	       	&81		&57			&77		&54		&--		&--  	\\
		Flashlight 	    &67		&53			&60 	&49 	&--	    &-- 	\\
		FreshBowser	    &90		&64			&52  	&31 	&65	    &41   	\\
		QiuShiBaiKe     &40 	&30 		&20     &14	    &--	    &-- 	\\
		SaoleiGame     	&78		&58		    &78  	&56 	&82	    &64    	\\
		SgSearch        &46     &38         &36     &27     &29     &18     \\
		Terminal 		&100  	&80			&100  	&80   	&100  	&76   	\\
		\hline\hline
	\end{tabular}
\end{table}

\begin{table}[!ht]
	\scriptsize
	\caption{Coverage Comparison on Source Code}\label{Experimental Applications 2}
	\centering
	\begin{tabular}{l|cc|cc|cc}
		\hline\hline
		\multirow{2}{*}{\textbf{App}} &\multicolumn{2}{c|}{\textbf{AppTag} }&\multicolumn{2}{c|}{\textbf{Monkey}} &\multicolumn{2}{c}{\textbf{Dynodroid}} \\
		& \textbf{MC} & \textbf{LC} & \textbf{MC} & \textbf{LC}  & \textbf{MC} & \textbf{LC}\\
		\hline
		aLogcat       &69	    &62 &57  &51 	&68		&60 	\\
		Budget	      &64		&56	&50	 &45	&--		&--		\\
		HotDeath	  &71	    &55 &54	 &43	&69	 	&53	    \\
		TippyTipper   &70 	    &64	&67  &59  	&-- 	&-- 	\\
		WhoHasMS &65	  &53	&60  &47	&--  	&--	    \\
		\hline\hline
	\end{tabular}
\end{table}

We compare the coverage of the test suites generated by different testing tools and the detailed information about coverage results are in Table \ref{Experimental Applications 1} and \ref{Experimental Applications 2} \footnote{Table \ref{Experimental Applications 2} only gives the results of five apps of the ten open-source apps, as \emph{EMMA} crashes due to engineering reasons and fails to measure the coverage of the rest five apps in our experiments. }. The first table shows that the \#CC (\%) and \#MC (\%) of all apps and the second one shows the \#MC (\%) and \#LC (\%) of the apps with source code. 
Besides, for some of the instances, Dynodroid fails to report the test results and we use ``--'' to represent them in the tables.
As shown in the tables, AppTag can reach higher coverage (about 20\% improvement on average) in most cases than Monkey and Dynodroid.

Let us use a specific app, i.e., \textit{Tippy Tipper} to demonstrate how the widget status and the back stack influence the GUI model of the app.
The app \emph{Tippy Tipper} is a popular open source calculator app, which can be used for calculating the tip amount for a meal. The user can enter the meal amount on the entry screen and get the result by clicking the \texttt{Calculate} button. Both the entry and result screen are attached with a menu. If the menu item \texttt{Setting} is selected, the app will take the user to the Setting window. The paths that can reach the Setting window are $\langle$Entry, Setting$\rangle$ and $\langle$Entry, Result, Setting$\rangle$. Although the final GUI window ``Setting'' they reach is the same, their back stacks are different. If we send a Back event to the app at these two windows, their behavior will be different. Therefore, they should not be merged as one state.
Besides, there is a \texttt{CheckBox} called \texttt{Enable Exclude Tax Rate} on the Setting window. If its status is ``checked'', the button \texttt{Tax Rate to Exclude} below it will be enabled on current Activity, or else disabled. Clicking the button \texttt{Tax Rate to Exclude} will drive the app to a new dialog window.
In this occasion, the status change of a widget influences other widgets related to it, furthermore, it influences the corresponding events of current state.
We measured the size of model influenced by widget status and the back stack in this case. Without considering these characteristics , the model contains 10 states and 172 transitions. And it will grow to 23 states and 450 transitions if these details are considered.


\begin{figure}[!ht]
	\centering
	\begin{tabular}{ll}
		\includegraphics[width=0.22\textwidth]{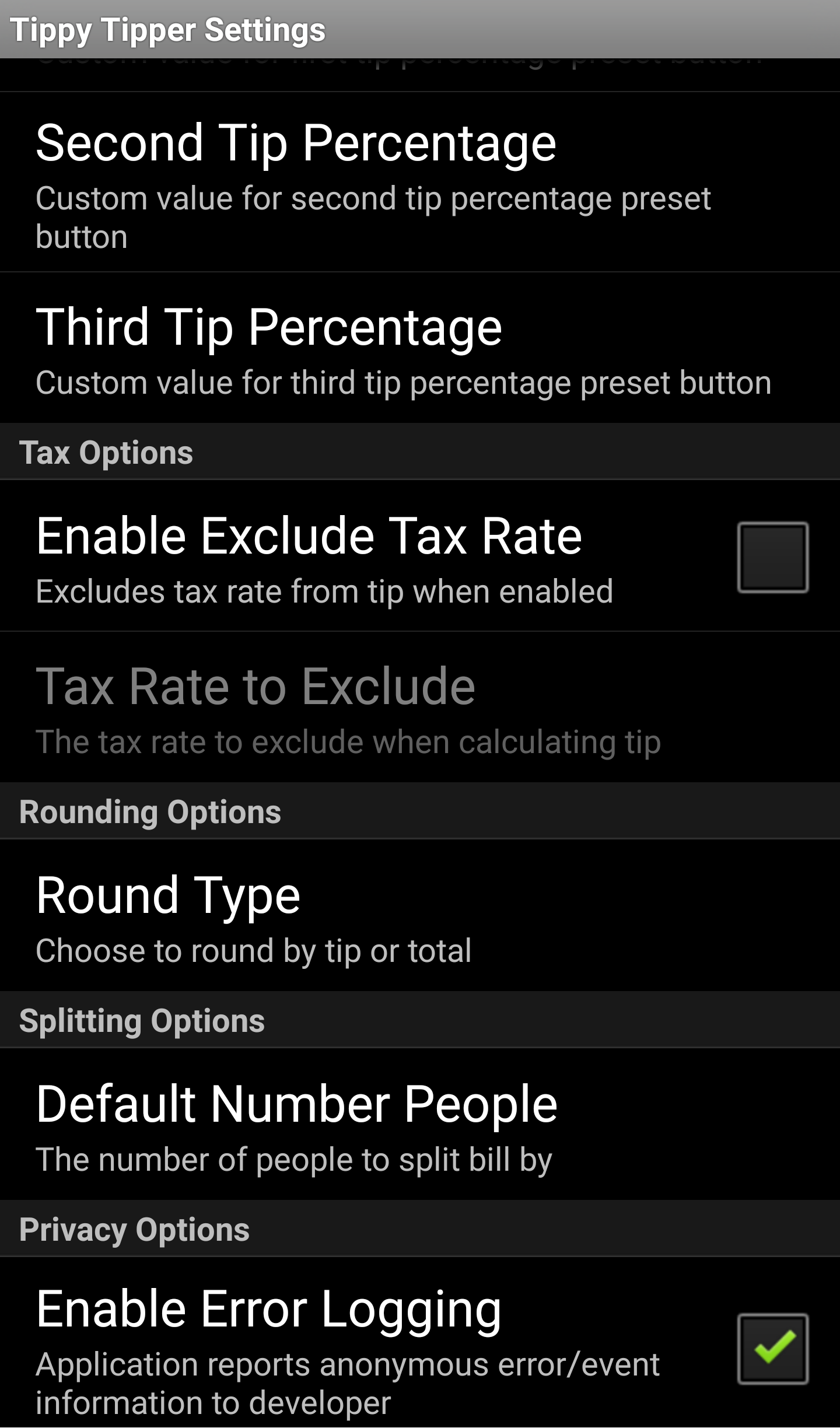}&\includegraphics[width=0.22\textwidth]{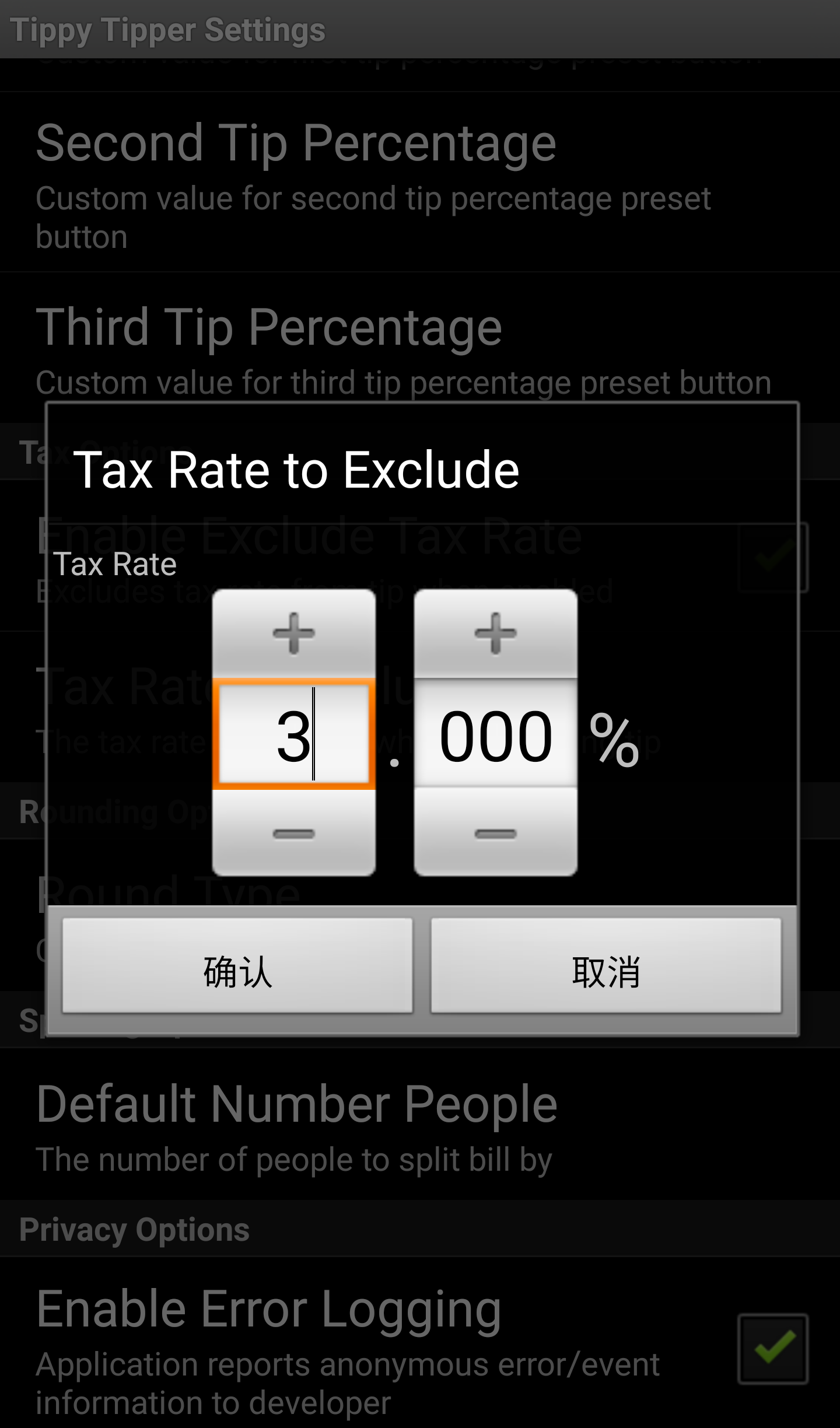}\\
	\end{tabular}
	\caption{Tippy Tipper Application}\label{Example2}
\end{figure}

\subsection{State Similarity and Model Size}\label{similarity}
In this subsection, we will discuss the impact of $S_T$ on the size of the model and the benefit to the coverage from high $S_T$.
The value of threshold $S_T$ is set from 0 to 1. We apply our tool to ten experimental instances to generate the LATTE model, and compare the numbers of states and transitions in the generated LATTE model and the method coverage of this model under different values of $S_T$. Here, the method coverage is calculated as the total number of methods in the app divided by the number of methods executed during the model construction, which is an indicator about the code coverage of the model. A high threshold $S_T$ may cause an extremely large even infinite model size. For example, apps with the file picking functionality will have an extremely large model size that causes its window to dynamically change a lot. Therefore, we set 3 hours as the upper bound of the execution time, record the final number of transitions and do not show the coverage result if the model construction is not finished within this bound.

\begin{figure}[!ht]
	\small
	\centering
	\begin{tabular}{c}
		\includegraphics[width=0.42\textwidth]{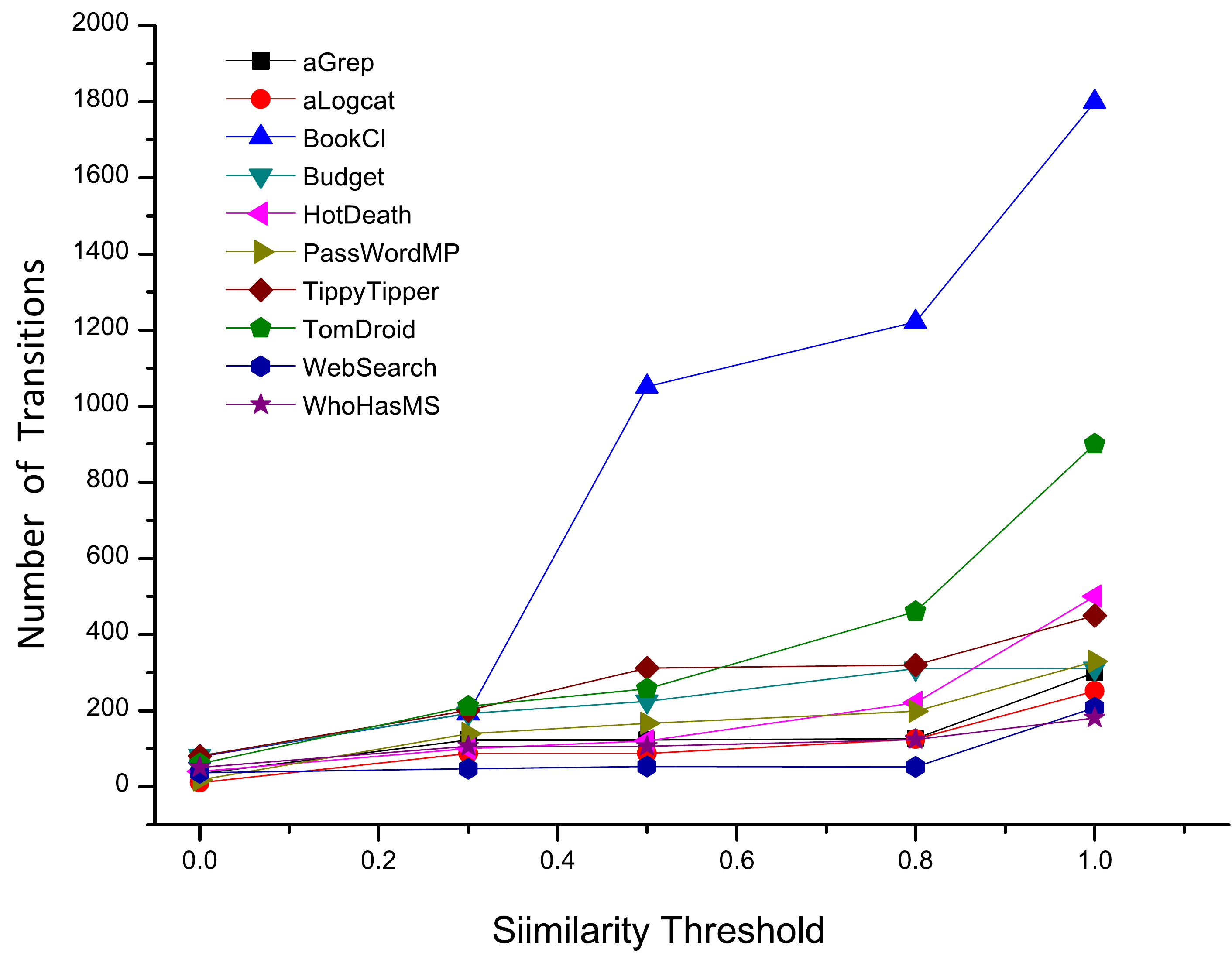}  \\
		(a) $S_T$ and Model Size  \\
		\includegraphics[width=0.42\textwidth]{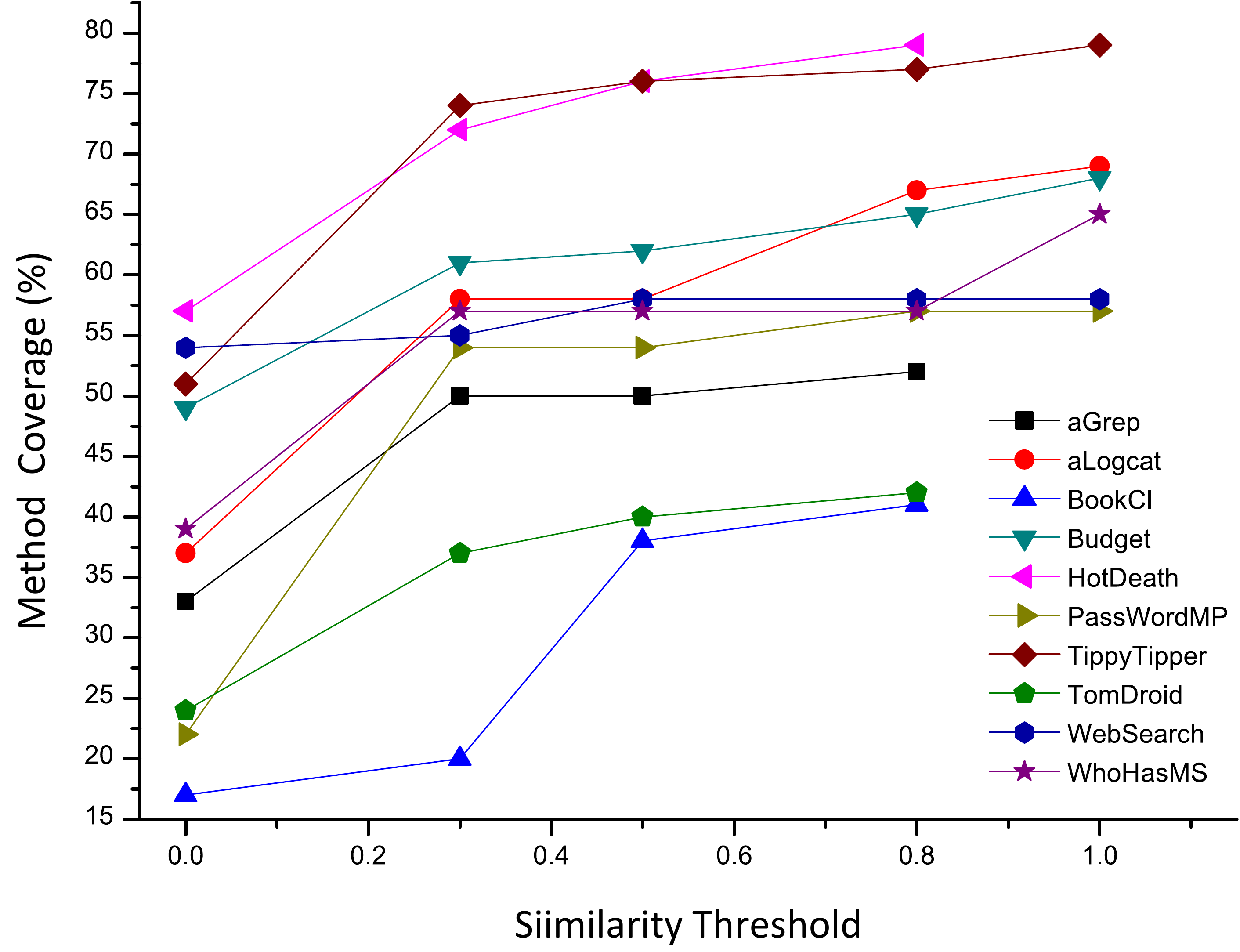}\\
		(b) $S_T$ and Coverage\\
	\end{tabular}
	\caption{Impact of Similarity Threshold}\label{Impact of Similarity Threshold}
\end{figure}

Fig. \ref{Impact of Similarity Threshold} demonstrates the tendency of the number of transitions in model and the method coverage on byte-code of generated test cases under different $S_T$ value for the ten open-source apps.
As we can see, with the increase of similarity threshold $S_T$, the size of the model increases dramatically. Obviously, the cost of model construction will increase accordingly. However, the coverage of test suites generated from the model will not increase significantly when $S_T$ reaches a certain level. Therefore, $S_T$ is a proper control variable to make a trade-off between the accuracy and efficiency. We find that 0.8 is a reasonable choice according to the result and use it as the default value in the following parts.

\subsection{Test Generation for Covering Targets}
In this part, we will give some targets including specific methods and Android APIs, and compare the event sequences generated by AppTag and Monkey.

\textbf{User-Developed Method Target.}
The experimental result of user-developed method target is shown in Table \ref{Method Target}. The second column lists the information of target, it can be a single label, or a set of several labels. The third column gives the minimal number of events for Monkey to cover the target. Note that the number of events $n$ denotes Monkey can cover the target by an event sequence whose number is in the range of $(n-1000,n]$. The last column presents the number of events that AppTag needed to cover the target. It is composed by two numbers, the number of events used for model construction and the length of sequence generated to cover the target. It shows that Monkey will trigger the target with a long sequence, especially when the AUT is complex or the target has a complicated execution logic, while AppTag can reach the target using an extremely short event sequence.

\begin{table}[!h]
\caption{user-Developed Method Target}\label{Method Target}
\centering
\begin{tabular}{|c|c|cc|}
\hline
\multirow{2}{*}{Application} & \multirow{2}{*}{Target} & \multicolumn{2}{c|}{Number of Events}  \\
            &        &  Monkey     &AppTag \\
\hline
HotDeath    &\tabincell{c}{showCardHelp} & 3000     & 241/3\\
\hline
PasswordMaker&\tabincell{c}{onExportClick}& 12000   &305/2\\
\hline
TomDroid    &\tabincell{c}{deleteNote,\\undeleteNote}&50000 &255/12\\
\hline
\tabincell{c}{WhoHas-\\MyStuff}&\tabincell{c}{updateDate,\\updateReturnDate}&  9000     &102/8  \\
\hline
Cradio      &\tabincell{c}{searchRadio,\\ addFavorite,\\addDelFavo}&4000&  394/14\\
\hline
SougouSearch&\tabincell{c}{addCard}&21000      &477/4 \\
\hline
\end{tabular}
\end{table}

\textbf{System API Target.}
Table \ref{API Target} shows the experimental result of system API target. The second column gives the system APIs we picked, including resource (VelocityTracker, AudioRecord, URL) and privacy (ContactsContract) related APIs. These experiments also demonstrate that AppTag can cover the target with a short sequence.

\begin{table}[!h]
\caption{System API Target}\label{API Target}
\centering
\begin{tabular}{|c|c|cc|}
\hline
\multirow{2}{*}{Application} & \multirow{2}{*}{Target} & \multicolumn{2}{c|}{Number of Events}  \\
   &           &  Monkey   &AppTag \\
\hline
AnyCut      &\tabincell{c}{ContactsContract\\(CONTENT\_URI)}&1000   &55/2\\
\hline
Budget      &\tabincell{c}{VelocityTracker\\(obtain)}&4000   &246/2\\
\hline
Voicesmith  &\tabincell{c}{AudioRecord\\(release)}& 3000   &135/2\\
\hline
\tabincell{c}{WhoHas-\\MyStuff}  &\tabincell{c}{ContactsContract\\(CONTENT\_URI)}&2000 &102/2\\
\hline
BubeiListen &\tabincell{c}{URL\\(openConnection)}&1000      &540/5\\
\hline
\end{tabular}
\end{table}

\section{Related Work}\label{related}

There are many kinds of test generation approaches on Android apps, including random testing, model-based and systematic testing. In this section, we will introduce several representative works based on these approaches and highlight the differences between our work and theirs.

\textbf{Random Testing. }
Monkey \cite{Monkey} is a widely used black-box testing tool, which can send sequences of random events to android apps. It is simple and fully automatic that can generate a great deal of test events within a short time. There are works based on Monkey for detecting GUI bugs \cite{DBLP:conf/icse/HuN11} and security bugs \cite{DBLP:conf/icse/MahmoodEKMMS12}. However, Monkey is not suitable for generating highly specific event sequences.

Dynodroid \cite{DBLP:conf/sigsoft/MachiryTN13} proposed by Machiry et al. provides a more efficient random GUI exploration approach compared with Monkey. They define several strategies for selecting events to guide the test generation procedure and support system event generation by instrumenting the Android framework.

\textbf{Model-based Testing. }
Model-based testing has been widely studied and applied in testing Android apps recently. A key point of model-based testing is to construct a model that can accurately depict the software behaviour.

Several researches construct the model by static analysis. W. Yang et al. \cite{DBLP:conf/fase/YangPX13} model the GUI behavior of application as a FSM, they proposed an approach that uses static analysis on Java source code of Android to extract actions associated with view components on a GUI state, and implemented a tool called ORBIT. S. Yang et al. \cite{DBLP:conf/kbse/YangZWWYR15} provided a similar model called Window Transition Graph (WTG), with a more accurate static callback analysis, it gives more careful model of currently-active windows stack and window transition. There are some differences between them and our work. The first one is that their model construction relies on the source code of the AUT, while we can handle the apk file directly. The second one is that they build a model statically that misses the changes of GUI screen during runtime.

Some researchers leverage dynamic analysis techniques to construct the model of the AUT. Amalfitano et al. \cite{Amalfitano12ASE} implemented a tool called AnroidRipper to explore the GUI views of the AUT. However, the model produced by this tool does not distinguish the different statuses of views in the same Activity. As a result, each GUI object may be included multiple times so that the size of the model is too large. Azim et al. \cite{DBLP:conf/oopsla/AzimN13} proposed Activity Transfer Graph (ATG) as their exploration model. They design a static analysis algorithm on the AUT to extract the Static Activity Transfer Graph (SATG), and use dynamic GUI exploration to handle dynamic activities layouts to complement the SATG. However, they regard the Activity as the minimum unit in ATG and also do not consider the different statuses of views and the back stack of the same Activity.

Model-based testing is also used to detect some specific bugs in Android apps, for example, Zhang et.al \cite{16AST} proposed a model-based test generation approach to expose resource leak defects in Android apps. They construct the WTG model proposed by Yang et al. \cite{DBLP:conf/kbse/YangZWWYR15} to describe the AUT and try to generate test cases for two important categories of neutral sequences, based on common leak patterns specific to Android. The main idea of their work is a bit like our work in that the common resource leak patterns can be regarded as the target. However, as we mentioned above, the WTG model is constructed by static analysis that may fail to describe the change of GUI screen during runtime. In addition, their target is only related to the GUI views rather than the code snippets that user concerned in our work.

\textbf{Systematic Testing. }
ACTEve \cite{DBLP:conf/sigsoft/AnandNHY12} is a concolic-testing tool that generates sequences of events automatically and systematically. It symbolically tracks events from the generated point in the framework to the handled point in the app, thus both the framework and the AUT need to be instrumented. Besides, ACTEve can handle system events as well as UI events. The limitation is that it only alleviates but not avoid the path explosion problem, and it currently handles only tap event.

Jensen et al. \cite{DBLP:conf/issta/JensenPM13} provide another concolic-testing approach that aims at automatically finding event sequences that reach a given target line in the application code. This approach improves automated testing for Android applications that are not computationally heavy but may have complex user interaction patterns. However, the work of symbolic execution can only process integer but not String or other complex data type. In addition, this approach also need a model for test case generation and they build it manually.

Choudhary et al. \cite{DBLP:conf/kbse/ChoudharyGO15} conduct an empirical study on existing testing tools for Android apps. An interesting discovery is that Monkey and Dynodroid which are based on the random exploration strategy can reach higher coverage than other tools with more sophisticated strategies. The major reason of this phenomenon is that most of the app behaviors can be exercised by generating only UI events. Our LATTE model can accurately and comprehensively describe these UI events and the experimental results show that the test cases generated by it can achieve similar coverage compared with Monkey and Dynodroid using shorter event sequences.

\section{Conclusion}\label{conclusion}


We proposed a dynamic way to model the AUT and then generate test cases to cover the given targets. To describe the AUT accurately, we presented a LATTE model with back stack, view and event information for the model-based testing of Android apps. Different from other dynamic approaches, our model also represents some of the code information via the label mechanism that guides the test generation procedure to cover the code snippets user concerned quickly. We have evaluated the effectiveness of our approach on several real-world apps, the result shows that it achieves the same coverage as the state-of-art tools, with a shorter event sequence.

We believe that our approach can greatly promote the effectiveness of Android testers and help developers for target directed testing. There are some possible ways to improve our approach. The cross application invocation is commonly used by developers and it influences the change of back stack according to its launch mode. Another potential improvement lies in the target set. Currently we regard a target as an unordered set of labels. If we introduce temporal logic to the target set, our work can be extended to dynamically check for some bugs with temporal properties. All these will be left as our future work.

\bibliographystyle{abbrv}
\bibliography{bib/ref}
\end{document}